\documentstyle[preprint,prd,aps,eqsecnum]{revtex}
\tighten
\begin{document}
\draft
\preprint{gr-qc/9405016\qquad NSF-ITP-94-45}
\begin{title}
	{\sf Behavior of Friedmann-Robertson-Walker Cosmological Models
			in Scalar-Tensor Gravity}
\end{title}
\author			{\sc Shawn J.~Kolitch
	\footnote{E-Mail address: \tt kolitch@nsfitp.itp.ucsb.edu}\\}
\address{Department of Physics\\University of California\\
		Santa Barbara, CA 93106-9530}

\author			{\sc Douglas M.~Eardley
	\footnote{E-Mail address: \tt doug@sbitp.ucsb.edu}\\}
\address{Institute for Theoretical Physics\\University of California\\
		Santa Barbara, CA 93106-4030}

\date{June~~~, 1994}

\maketitle
\begin{abstract}
We analyze solutions to Friedmann-Robertson-Walker cosmologies in Brans-Dicke
theory, where a scalar field is coupled to gravity.  Matter is modelled by
a $\gamma$-law perfect fluid, including false-vacuum energy as a special case.
Through a change of variables,
we reduce the field equations from fourth order to second order, and they
become equivalent to a two-dimensional dynamical system.  We then analyze
the entire solution space of this dynamical system, and find that many
qualitative features of these cosmologies can be gleaned, including standard
non-inflationary or extended inflationary expansion, but also including
bifurcations of stable or unstable expansion
or contraction, noninflationary vacuum-energy dominated models, and several
varieties of ``coasting," ``bouncing," ``hesitating," and ``vacillating"
universes.  It is shown that inflationary dogma, which states that a universe
with curvature and dominated by inflationary matter will always approach a
corresponding flat-space solution at late times, does not hold in general for
the scalar-tensor theory, but rather that the occurence of inflation depends
upon the initial energy of the scalar field relative to the expansion
rate.  In the case of flat space ($k=0$), the dynamical system formalism
generates some previously known exact power-law solutions.
\end{abstract}
\pacs{04.50.+h, 98.80.Hw, 02.30.Hq}

\def\fig#1{Fig.~{#1}}			% For figure numbers
\def\Equation#1{Equation~\(#1)}		% For citation of equation numbers
\def\Equations#1{Equations~\(#1)}	%	ditto
\def\Eq#1{Eq.~\(#1)}			%	ditto
\def\Eqs#1{Eqs.~\(#1)}			%	ditto
\let\eq=\Eq\let\eqs=\Eqs		%	ditto
\def\(#1){(\call{#1})}
\def\call#1{{#1}}
\def\square{\kern1pt\vbox{\hrule height 1.2pt\hbox{\vrule width 1.2pt\hskip 3pt
   \vbox{\vskip 6pt}\hskip 3pt\vrule width 0.6pt}\hrule height 0.6pt}\kern1pt}

\section{Introduction}
\label{intro}

It was first recognized by Guth that inflation, {\it i.e.,\ } a period
of superluminal
expansion in the early universe, solves many of the problems of the standard
model of cosmology, a.k.a{.} the ``hot big bang'' theory\cite{Guth}.  These
include the smoothness/horizon problem, the flatness/age problem, the
structure problem and the massive relics problem.  (A review of these problems,
and their resolution in an inflationary scenario, is given in
\cite{Kolb}.)
However, in the original scenario proposed by Guth, one finds that the
deSitter expansion induced by vacuum energy density (or, equivalently, by
a cosmological constant) in ordinary general relativity is impossible to
exit from via barrier penetration in a way that permits sufficient inflation
to solve the aforementioned problems, while still reheating
the universe afterwards\cite{Wein}.  In particular, it can be shown that if the
tunneling transition rate for the inflaton field is small enough to allow for
sufficient inflation, then it is too small to nucleate true-vacuum
bubbles fast enough to collide and reheat the universe, and the phase
transition to a radiation-dominated era is never completed.  A number
of possible solutions have been proposed for this ``graceful exit problem,"
including an extremely flat inflation potential (``new inflation"),
so that inflation occurs as the field ``rolls down'' towards its true-vacuum
state, without any barrier penetration being necessary\cite{Linde}; and
including ``chaotic inflation," in which only certain regions of the
universe inflate to large size before the true vacuum is restored\cite{Linde2}.

One such proposal,
``extended inflation''\cite{La}, based on the Brans-Dicke (BD) theory of
gravity \cite{BD}, solves the graceful exit problem while still allowing
the phase transition that ends inflation to be completed via bubble
nucleation, by employing power-law rather than exponential expansion.
This permits eventual complete reheating even if the tunneling transition rate
for the inflaton is small enough to ensure that inflation continues at least
until the relevant cosmological problems are solved.

In its original form, this proposal fails due to the creation of
unacceptably large microwave background perturbations by collisions
between big bubbles\cite{EJW}.  To avoid this ``big bubble problem",
the coupling constant must take on a value in the range $\omega \le 25$
during inflation, in conflict with limits set by recent experiments of
$\omega > 500$ \cite{RDR}.  This conflict may be circumvented by further
alterations of the gravitational sector, such as a potential for the
BD scalar\cite{PJS}.  In addition to the big bubble problem,
however, one finds that the spectrum of density perturbations produced
by quantum fluctuations of the BD scalar during inflation is
distorted away from scale-invariance, so that COBE measurements of the
spectral index of the perturbations put a further constraint on the
allowed range of $\omega$ during inflation.  Taken together, these
constraints appear to rule out many extended inflation models\cite{LL}.
Models which survive include those with a variable coupling constant such
as hyperextended inflation\cite{S-A},  and hybrid models which include
both a first-order phase transition and a period of slow-roll\cite{RH}.

Recent interest in BD theory is also motivated by its relation
to low-energy bosonic string theory.  We can write the action for BD
theory as
\begin{equation}
	S_{BD} = \int d^4x \sqrt{-g} \left(-\phi R + \omega {{\phi^{,\mu}
		\phi_{,\mu}}\over {\phi}} + 16\pi
		{\cal \char'114}_m \right),
\label{BDaction}
\end{equation}
where $R$ is the scalar curvature, $\omega$ is the dimensionless BD
coupling constant, and $\phi$ is the BD scalar field.  The low-energy
effective action for bosonic string theory can be written in the
form\cite{String}
\begin{equation}
	S_{eff} = \int d^4x \sqrt{-g}e^{-2\Phi}\left(R + 4{\Phi^{,\mu}
		\Phi_{,\mu}}-{1 \over 12}H^2 \right).
\label{EFFaction}
\end{equation}
Here $R$ is again the scalar curvature, but now $\Phi$ is the dilaton field
and $H^2 \equiv H_{\mu \nu \rho} H^{\mu \nu \rho}$, where $H_{\mu \nu \rho}$
is the totally antisymmetric three-form field.  If we make the substitution
$\phi = e^{-2\Phi}$ we see that, considering only the scalar and tensor
components and neglecting the matter, the two theories
are identical in the low-energy limit if $\omega = -1$.  It should be
emphasized, however, that the two theories differ in their couplings of the
scalar field to the other matter.
The renewed interest in scalar-tensor gravitation has led to several
recent investigations into the generation of exact solutions for cosmology
in such theories\cite{Barrow93}, as well as to some qualitative studies of
the models which result\cite{Barrow90,Damour93}.

There is considerable interest, therefore, both from inflationary
cosmology and from string theory, to revisit cosmology in the
scalar-tensor gravity of Brans and Dicke.  Many papers were of course
written on BD cosmology in the 1960's and 1970's\cite{Green,Nariai,%
Morganstern,TV,bounce};
interest then declined, partly because observational tests of gravity put
increasingly stringent lower limits on the BD $\omega$ in the present
universe, and partly because fundamental scalar fields fell out of
theoretical favor.  However, string theory restored the theoretical
credibility of the scalar field, because of the necessary appearance of
the dilaton, leading to the realization that the scalar field can, and
indeed must, gain a mass and decouple from present-day long-range
physics.  So, one's point of view on BD cosmology has changed:
it does not apply at the present time, but may apply during eras of
inflation, or even of quantum gravity, in the early universe.
Differences from general relativity may be relevant to the viability of
string theory at energies below the Planck scale, and to inflationary
scenarios which incorporate scalar fields coupled to the other matter.

However, if one is to use the scalar-tensor theory in the place of general
relativity in early cosmology, a natural and necessary first step is the
delineation of the full spectrum of cosmological models which are possible
in the theory.  With the addition of an adjustable coupling constant
and a scalar field requiring two initial conditions, it should come as
no surprise that a much richer variety of qualitatively distinct
models are available here than in general relativity.  The purpose
of this paper is to consolidate these models into a complete and coherent
representation of their character.  To our knowledge there has been no
previous attempt to do this.  An essential practical tool here is a new
transformation of variables that reduces the system of equations from fourth
order to second order, greatly simplifying the analysis.

After this work was completed, we learned from D.~Wands that
there is substantial overlap between our work and his Ph.D.\ thesis
\cite{Wands}, including the reduction of the equations to a second order
system.

The paper is organized as follows:  after writing down the equations
for homogeneous and isotropic cosmology in the scalar-tensor theory in
Sec.~II, we reduce those equations to a 2-dimensional dynamical system
in Sec.~III.  The analysis of the dynamical system, which
contains descriptions of all of the qualitatively distinct models
allowed by the theory, is performed in Sec.~IV.  In Sec.~V we use
the dynamical system formalism to rederive some previously known
exact power-law solutions to the theory in flat space, and Sec.~VI
presents conclusions.

\section{Cosmology in BD Theory}
\label{BDcosmo}

The field equations for BD theory can be found by varying the action,
Eq.~\(\ref{BDaction}), with respect to the metric and the scalar field.
They are
\begin{equation}
	2\omega \phi^{-1} \square \phi - \omega {{\phi^{,\mu}
		\phi_{,\mu}} \over {\phi^2}} + R = 0
\label{BDphieq}
\end{equation}
and
\begin{equation}
	R_{\mu \nu} - {1\over 2}g_{\mu \nu}R = 8\pi \phi^{-1} T_{\mu \nu} +
		\left({\omega \over \phi^2} \right)(\phi_{,\mu} \phi_{,\nu}
 		- {1\over 2}g_{\mu \nu} \phi^{,\rho} \phi_{,\rho})
		+ \phi^{-1}(\phi_{,\mu {;}\nu} - g_{\mu \nu}
		\square \phi),
\label{BDgeq}
\end{equation}
where
\begin{equation}
	\square \phi \equiv {1 \over \sqrt{-g}}{\partial \over
		{\partial x_\mu}}\left(\sqrt{-g}{{\partial \phi}
		\over {\partial x^\mu}}\right).
\label{boxdef}
\end{equation}
Taking the trace of Eq.~\(\ref{BDgeq}) and combining the result with
Eq.~\(\ref{BDphieq}), we find
\begin{equation}
	\square \phi = \left({8\pi} \over {3 + 2\omega} \right)
		T^\mu {}_\mu.
\label{BDboxeq}
\end{equation}
Eqs.~\(\ref{BDgeq},\,\ref{BDboxeq}) may be viewed as the
fundamental equations for BD
theory.  We take as the background metric the homogeneous and isotropic
Friedmann-Robertson-Walker (FRW) line element, which may be written in
comoving coordinates as
\begin{equation}
	ds^2 = -dt^2 + a^2 (t) \left[{{dr^2} \over {1-kr^2}} + r^2 d\Omega^2_2
		\right],
\label{FRWmetric}
\end{equation}
where $a(t)$ is the scale factor and $k = 0, \pm 1$ is chosen to represent
the universe as a space of constant zero, positive or negative curvature,
respectively.  Then we find that the nontrivial components of the field
equations, Eqs.~\(\ref{BDgeq},\,\ref{BDboxeq}), are
\begin{eqnarray}
	\left({\dot a \over a} + {\dot \phi \over {2\phi}} \right)^2
		+ {k \over a^2} =&& \left({2\omega + 3} \over 12 \right)
		\left({\dot \phi \over \phi} \right)^2 +
		{8\pi \rho \over {3\phi}}	\label{BDFRW1}		\\
	-{1 \over {a^3}}{d \over dt}(\dot \phi a^3) =&&
		\left({8\pi} \over {3 + 2\omega} \right)T^\mu{}_\mu,
\label{BDFRW2}
\end{eqnarray}
where we have written only the time component of the metric field equation,
since the spatial components are dependent on it through the Bianchi
identities.  Assuming a perfect fluid form for the stress-energy tensor,
{\it i.e.,\ } $T_{\mu \nu} = diag(\rho, p, p, p)$, we also find the
usual conservation
equation (the time component of $T^{\mu \nu}{}_{;\nu} = 0$):
\begin{equation}
	\dot \rho = -3 {\dot a \over a} (p + \rho).
\label{BDFRW3}
\end{equation}

Eqs.\ (\ref{BDFRW1}-\ref{BDFRW3}) are the equations for FRW cosmology in
BD theory.  Note that we have a system of one second-order and
two first-order differential equations, so that we need four pieces of
initial data to completely specify a solution in this theory, as well as
an equation of state $p(\rho)$ and values of $\omega$ and $k$.  By
contrast, when solving for the Friedmann models in general relativity, we
have two first-order equations and hence need only two pieces of initial
data.  The difference stems from the fact that in BD theory,
$\phi^{-1}$ acts like a time-varying gravitational constant obeying the
second-order Eq.~\(\ref{BDFRW2}), so that we need two additional initial
conditions in order to close the system.  An interesting consequence of
this is that even after specifying initial conditions for $a$, $\phi$ and
$\rho$, we are left with a one-parameter family of solutions parametrized
by the initial value of the rate of change of the scalar field.  Indeed,
as has been noted by other authors\cite{Green,SWein}, the character of
the solutions at early times depends solely upon the value of the
constant $a^3\dot\phi$ at the initial singularity.  At later times,
however, most solutions become indistinguishable from those where
$a^3\dot\phi=0$ at the singularity.  Accordingly, there has been a
concentration in the literature on the three-parameter family of
solutions characterized by $a^3(0)\dot\phi(0)=0$.  However, one must bear
in mind that there has been a loss of generality when this boundary
condition is imposed.

We turn now to theoretical considerations regarding the possible values
of the coupling constant $\omega$.  First, inspection of Eq.~\(\ref{BDFRW1})
reveals that at early times, the curvature term vanishes relative to the
other terms provided only that $\ddot a < 0$ as $a\to 0$, {\it i.e.,\ }
provided that the initial epoch in the history of the universe was not
inflationary.  If we make the further, seemingly reasonable assumptions
that $\rho > 0$ and $\phi > 0$, then we see that we must have ${\omega\ge
-3/2}$ in order to satisfy Eq.~\(\ref{BDFRW1}) for all times.  Perhaps a
more compelling argument for this constraint on $\omega$ arises when
BD theory is reformulated in a way that preserves the form of
the Einstein equations (the so-called ``Einstein frame''), by treating
the scalar as a dynamical field which couples to all of the other
matter\cite{Dicke}.  When this is done, one finds that the stress-energy
for the scalar field is
\begin{equation}
	T_{\mu\nu}^{(\phi)} = {{2\omega+3}\over{16\pi G_0\phi^2}}
		(\phi_{,\mu}\phi_{,\nu}-{1\over2}g_{\mu\nu}
		\phi_{,\rho}\phi^{,\rho}),
\label{Tscalar}
\end{equation}
so that assuming homogeneity,
\begin{equation}
	T_{00}^{(\phi)}=T_{ii}^{(\phi)}={{2\omega+3}\over{32\pi G_0}}
		{\left({\dot\phi}\over{\phi}\right)}^2.
\label{Tscal2}
\end{equation}
{}From this, we see that the energy density of the scalar will be
nonnegative in the Einstein frame only for ${\omega\ge -3/2;}$  thus, for
example, one can only prove the existence of black holes or of a
cosmological singularity for $\omega$ in this range.  As a result of
these considerations, we take ${\omega\ge -3/2}$ in what follows.

\section{Reduction of the Field Equations to a Dynamical System}
\label{DSred}

In this section, we change variables in order to transform the
fourth-order system specified by Eqs.~ \(\ref{BDFRW1}-\ref{BDFRW3}) into a pair
of coupled second-order equations in which, however, only first
derivatives of the new variables appear.  The resulting equations may
then be analyzed using the established tools of dynamical systems theory.
First we switch to conformal time $d\tau = dt / a(t)$, and define the new
variables
\begin{eqnarray}
	\beta \equiv&& \left({a' \over a} + {\phi ' \over {2\phi}} \right)
		\label{betadef}						\\
	\sigma \equiv&& \left({{2\omega + 3} \over 12} \right){\phi' \over
		\phi} \equiv A {\phi' \over \phi},	\label{sigdef}
\end{eqnarray}
where primes represent derivatives with respect to conformal time, and
\begin{equation}
	A \equiv \left({{2\omega+3}\over 12}\right).
\label{Adef}
\end{equation}
Next, we parametrize the equation of state by writing
\begin{equation}
	p = (\gamma - 1) \rho.
\label{pdef}
\end{equation}
Since the three-form field is necessarily present in the low energy
limit of string theory, it is of interest to consider how it might bear
on the present study.  If we assume a homogeneous and isotropic universe,
then it follows that
\begin{eqnarray}
	H_{0\mu\nu} =&& 0, 	\nonumber 			\\
	H_{123} =&& h(t).
\end{eqnarray}
If we then use the fact that
\begin{equation}
	T_{\alpha\beta} = {{-2\over\sqrt{-g}} {{\delta(\sqrt{-g}
		{\cal\char'114})}
		\over {\delta g^{\alpha\beta}}}}
\end{equation}
to calculate the stress-energy for the three-form under these assumptions, we
find that it can be modelled phenomenologically as a perfect fluid with
equation of state $p=\rho$.  Then we have, for example,
\begin{equation}
	\gamma = \cases{  0,   & false-vacuum energy; \cr
			  1,   & pressureless dust; \cr
			4/3,   & radiation; \cr
			  2,   & 3-form field. \cr}
\end{equation}
Now we can rewrite Eqs.\ (\ref{BDFRW1}-\ref{BDFRW3}) as
\begin{eqnarray}
	\beta^2 + k =&& {{8\pi \rho a^2} \over {3\phi}}+{\sigma^2 \over A}
			\label{BDred1}					\\
	{1\over a^2} (a^2 \phi')'=&& (1-3\gamma /4){{8\pi \rho a^2}\over {3A}}
			\label{BDred2}					\\
	\rho' = -3\gamma {a' \over a} \rho =&& -3\gamma \rho
		\left(\beta - {\sigma \over {2A}}\right) \label{BDred3}
\end{eqnarray}

Note that although false-vacuum energy may play a role in the dynamics of
the theory, the cosmological constant $\Lambda$ does not explicitly
appear in the field equations as we have written them.  Unlike in general
relativity, where the inclusion of $\Lambda$ is exactly equivalent to the
presence of vacuum energy, in BD theory the two are
qualitatively distinct.  This is the case because whereas in GR the
vacuum energy enters the field equations through the term $8\pi G \rho_v$
(which one can equate to $\Lambda$), in BD theory it enters
through the term $8\pi \rho_v / \phi$, clearly not in general a constant.
Others have considered the case $\Lambda \ne 0$ with no matter
present\cite{Romero}; in a separate paper, the present authors will
consider the system with both $\Lambda \ne 0$ and additional matter
present\cite{Kolitch}.

Combining Eq.~\(\ref{BDred1}) with Eq.~\(\ref{BDred2}) and using
the definitions of $\beta$ and $\sigma$, we can derive the expression
\begin{equation}
	\sigma' = (1-3\gamma /4)(\beta^2 + k - \sigma^2 /A) - 2\beta \sigma.
\label{DS1}
\end{equation}
Next we take the conformal time derivative of Eq.~\(\ref{BDred1}), and
use this in
conjunction with Eq.~\(\ref{BDred3}) and Eq.~\(\ref{DS1}) to obtain
\begin{equation}
	\beta' = (1-3\gamma /2)(\beta^2 + k) - (3-3\gamma /2)\sigma^2 /A.
\label{DS2}
\end{equation}
Eqs.~\(\ref{DS1},\,\ref{DS2}) constitute a dynamical system (in conformal
time) for the variables $\beta$ and $\sigma$, and as such it is conducive
to both quantitative and qualitative analysis.

\section{Qualitative Analysis}
\label{qualanal}

Although it is certainly agreeable to have some exact solutions to the
system specified by Eqs.\ (\ref{BDFRW1}-\ref{BDFRW3}), such solutions typically
correspond to a particular choice of the initial data.  Furthermore, it
is difficult to synthesize the conglomeration of known particular
solutions to the theory into a picture which reveals its overall
character.  It is therefore desirable to obtain a more complete
understanding of the solution space, which may be accomplished through a
qualitative analysis of the dynamical system (\ref{DS1},\,\ref{DS2}) in the
$\beta$--$\sigma$ plane, accompanied by a ``translation'' of the results
back into the physically meaningful $a$--$\phi$ plane.  Before beginning
the analysis, however, we shall make some general comments regarding the
system.

First, let us consider what an equilibrium point of the system
represents.  Whenever \hbox{$\beta' = \sigma' = 0$,} this implies that
$\dot a = \beta - \sigma/2A$ is a constant, so that
\begin{equation}
	a(t) = \left(\beta - {\sigma \over {2A}}\right)t + a_0.
\label{EQa}
\end{equation}
Also it is straightforward to show that at such a point,
\begin{equation}
	\phi(t) = CA\left[(\beta-\sigma /2A)t + a_0\right]^
		{\sigma / [A(\beta - \sigma /2A)]}
\label{EQphi}
\end{equation}
unless $\beta = \sigma /2A$, in which case
\begin{equation}
	\phi(t) = \phi_0 \exp(\sigma t /Aa_0).
\label{EQphi2}
\end{equation}
Hence we see that a fixed point in the $\beta$--$\sigma$ system represents
either a solution where the scale factor changes linearly with time while
the scalar varies as a power of time, or a static universe with exponentially
varying scalar.

Next, it will be of interest, as we consider the various fixed points of
the system, to determine whether or not a given point lies in the physical
domain $\rho > 0$.  We can check this using Eq.~\(\ref{BDred1}), which reads
\begin{equation}
	\beta^2 + k = {\sigma^2} / A + {{8\pi \rho a^2} \over 3\phi},
	\eqnum{\ref{BDred1}}
\end{equation}
so that (assuming $\phi > 0$,) any point $(\beta_0,\sigma_0)$ satisfying
\begin{equation}
	{\beta_0}^2 + k > {{\sigma_0}^2} / A
\end{equation}
will lie in $\rho > 0$, whereas points satisfying
\begin{equation}
	{\beta_0}^2 + k < {{\sigma_0}^2} / A
\end{equation}
lie in $\rho < 0$ and thus do not represent physical solutions.

A brief discussion regarding equations of state is in order at this
point.  Of particular interest is the question of what equations of state
will lead to inflation in this theory.  As was mentioned in Section II,
BD theory can be reformulated through a conformal transformation so
as to preserve the form of Einstein's equations, while viewing the scalar
field simply as a part of the matter (albeit a peculiar part) which
couples dynamically to every other field, and hence to the particle masses.
Using this form of the theory, we can write the usual equations of general
relativity for homogeneous, isotropic cosmology, {\it i.e.,\ }
\begin{equation}
	{\left({\dot a}\over a \right)^2} + {k \over {a^2}} =
		{{8\pi G \rho_{tot}}\over 3},
\end{equation}
\begin{equation}
	{\ddot a \over a} = {{-4\pi G}\over 3}(\rho_{tot}+3p_{tot}),
\end{equation}
where the subscripts are a reminder that these quantities include
contributions from the scalar field in addition to the ``regular'' matter.
As we saw previously [cf. Eq.~\(\ref{Tscal2})], such contributions are given by
\begin{equation}
	p_{\phi} = \rho_{\phi} = {{2\omega +3}\over{32\pi G_0}}
		{\left({\dot\phi}\over
		\phi\right)}^2.
\end{equation}
We see that inflation will occur only when $p_{tot} < -\rho_{tot}/3$. For
models which are not dominated by the scalar field, this just reduces to
the usual condition for inflation in general relativistic cosmology, $p_m
< -\rho_m /3 \; (\gamma < 2/3)$.  It is also clear, however, that the
contribution of the scalar field to the stress-energy tends to act
against inflation (and, more generally, against expansion), so that
inflation may not occur if the scalar field dynamics become sufficiently
dominant, regardless of the equation of state of the other matter.

Finally, the applicability of the singularity theorems of Hawking and
Penrose\cite{Singthm} to BD theory is of interest.  In general
relativity, these theorems prove the existence of a cosmological
singularity for the Friedmann-Robertson-Walker models, provided that
Einstein's equations and the strong energy condition
are all satisfied.  For a perfect fluid, this latter condition reduces to
$\rho + 3p \ge 0$, {\it i.e.,\ } only for inflationary matter is there
the possibility
of a cosmological model without a singularity.  To apply the theorems
to BD theory, we must transform to the Einstein frame, where
the Einstein equations are formally preserved.  Under this transformation,
one finds that\cite{Dicke}
\begin{equation}
	a(t) \to G_0^{-1/2} \phi^{-1/2} a(t),
\end{equation}
from which it follows that
\font\small=cmr7 scaled\magstep1
\begin{equation}
	\left({\dot a \over a} + {\dot\phi\over{2\phi}}
		\right)_{\small\hbox{BD}} \to
		\left(\dot a \over a\right)_{\small\hbox{E}}.
\end{equation}
Hence we see that an expanding universe in the Einstein frame is anything
with (in our notation) $\beta > 0$ in the BD frame.  As a result
of this, a model with ordinary matter which appears to contract and then
smoothly reexpand
in the BD frame will nevertheless be consistent with the
singularity theorems so long as $\beta > 0$ during the transition, for then
the model is in fact always expanding in the Einstein frame and the
singularity in the past would be due to the infinite value of the scalar.

Now let us proceed with the analysis.  Rewriting the results of section III,
we start with
\begin{eqnarray}
	\beta' =&& \left(1-{3\gamma \over 2}\right)(\beta^2 + k)
		- \left(3-{3\gamma \over 2}\right)
		\left({\sigma^2 \over A}\right),   \eqnum{\ref{DS2}}  \\
	\sigma' =&& \left(1-{3\gamma \over 4}\right)
		\left(\beta^2 + k - {\sigma^2 \over A}\right)
		- 2\beta \sigma, \eqnum{\ref{DS1}}
\end{eqnarray}
where, as before,
\begin{eqnarray}
	\beta \equiv&& \left({{a'}\over a} + {{\phi'}\over{2\phi}}\right),
					\eqnum{\ref{betadef}}		\\
	\sigma \equiv&& \left({{2\omega +3}\over 12}\right)
		{{\phi'}\over{\phi}} \equiv A{{\phi'}\over{\phi}},
					\eqnum{\ref{sigdef}}
\end{eqnarray}
and we are in conformal time $d\tau = dt/a$.  It is convenient to divide
the equilibrium points $(\beta_0,\sigma_0)$ into two categories:
\begin{eqnarray}
	\underline{(i)} \quad &&{\beta_0}^2 + k = 0, \quad \sigma_0 = 0
		\qquad (k=0,-1)	\nonumber			\\
	\underline{(ii)} \quad &&\beta_0 = \pm \sqrt{k \over \mu}
		(1-3\gamma /4),	\quad \sigma_0 = \pm \sqrt{k \over \mu}
		A(1-3\gamma /2)	\qquad (\forall k)	\label{EQpts}
\end{eqnarray}
where
\begin{equation}
	\mu \equiv 3A(1-\gamma /2)(1-3\gamma /2) - (1-3\gamma /4)^2.
\end{equation}

First consider $k=0$, in which case all of the equilibrium points,
regardless of the values of $\gamma$ and $\omega$, take on the values
$(\beta_0,\sigma_0)=(0,0)$.  According to our recent discussion, this
point represents the static solution $a=a_0, \ \phi=\phi_0$.  Inspection
of the field equations also reveals that this solution is only valid in
the true vacuum $\rho=0$.  However, although the solution appears at
first glance to be unphysical, it may also be viewed as the usual,
asymptotically approached solution $\dot a \to 0, \ a \to \infty$ for a
flat, expanding universe.  The fact that the rate of change of the scalar
field approaches zero at large times is simply the statement that the
strength of the gravitational interaction becomes nearly constant, as it
must if the theory is to agree with what we observe in the universe
today\cite{gdot}.  \fig{1a} shows the numerically integrated solution
curves in the $\beta$--$\sigma$ plane for a flat, matter-dominated (p=0)
universe, where we have arbitrarily selected $A=1 \ (\omega = 9/2)$ for
the plot.  The shaded areas represent unphysical regions of negative
energy density, and the dark line represents the transition between
expansion and contraction, this transition being forbidden for positive
energy densities in the present case.  The solutions to the right of
$\beta = 0$ are initially expanding, and we see that they asymptotically
approach the equilibrium point $(\beta_0,\sigma_0)=(0,0)$, an
interpretation of which has already been given.  The solutions to the
left of $\beta=0$ are initially contracting, and will continue to do so
until a singularity is reached.  We see that the solutions approach and
depart from equilibrium along a particular path, which represents a known
power-law solution to the flat-space field equations.  We shall consider
such exact solutions, and their relation to the general class of
cosmological models, in Section~V.  For the time being, we simply mention
that the attractive nature of such solutions near equilibrium is a
general feature of the theory, and that these solutions typically
correspond to the particular choice of initial data $a^3 \dot\phi = 0$ at
the singularity.

A qualitative change occurs when the coupling constant takes on a value
in the range $-3/2 \le \omega \le 0$.  \fig{1b} is similar to \fig{1a}, except
that here we take $\omega = -1/2$.  Note that now the line representing
$\dot a = 0$ exists in the physical regime where $\rho > 0$.  In fact,
one sees that models exist where the universe is initially contracting,
but then makes a smooth transition to expansion without meeting a
singularity.  This behavior has been noted by other authors\cite{bounce},
and models with this feature are sometimes called ``bounce'' universes.
It should be mentioned that such models are present for $\omega = -1$,
the value corresponding to the low-energy limit of string theory.

Next we examine the flat space models with an inflationary equation of
state.  \fig{1c} shows the solution curves for $k=0$ and $p=-\rho$, where
again we arbitrarily select $\omega = 9/2$.  The attracting nature of
the particular solution represented by the line $\beta = 3\sigma$ is
quite apparent, and this is in fact the inflationary power-law
solution which has been found previously by other authors\cite{BDinf}.
Here all expanding universes asymptotically approach this state, and once
again all contracting universes must reach a singularity, although we
see that as they do so, they approach the true-vacuum solutions
(represented by the dark lines at the border of the shaded region).
These are examples of what may be termed `$\phi$-dominated' models;
{\it i.e.,\ } models where the scalar field
dynamics overwhelm the effect of the other matter and thus attract the
universe towards true-vacuum behavior.  Nevertheless, these models must
contract until a singularity is reached.

Again the character of the solutions changes in the range
$-3/2 \le \omega \le 0$,
in a manner similar to the noninflationary case.  Here, however, not only
do we see that there exist ``bounce'' models,
but we also find that inflation is no longer possible in this range of
$\omega$.  Rather, all models approach the equilibrium solution
$\dot a = 0, \ a \to \infty$ as they would for noninflationary equations
of state; another example of `$\phi$-domination'.  This behavior is
illustrated in \fig{1d}.  In the range $0<\omega \le 1/2$, inflation is
still not possible, although the ``bounce'' models are absent.

Now let us consider the case of the open universe, $k=-1$.  First
consider the equilibrium points in category (i), {\it i.e.,\ } $(\beta_0,
\sigma_0)=(\pm 1,0)$.  Here the corresponding solutions are \hbox{$a(t) =
a_0 \pm t,\ \phi(t) = \phi_0$} which are, again by inspection of the
field equations, valid only for $\rho = 0$.  The physical interpretation
of the point $(\beta_0,\sigma_0)=(+1,0)$ is straightforward; we can view
it either as representing a particular vacuum solution, or as the
asymptotically approached endpoint of an open, expanding universe.  The
contracting solution represented by the point $(\beta_0,\sigma_0)=(-1,0)$
does not lend itself to interpretation quite so easily; to aid us in our
efforts we can perform a stability analysis.  This proceeds by
linearization, {\it i.e.,\ } we check the eigenvalues of the Jacobian
matrix for the system $\beta ' = f(\beta, \sigma); \ \sigma ' = g(\beta,
\sigma)$.  In cases where the eigenvalues of the Jacobian all have
nonvanishing real part, the fixed point is called hyperbolic and we can
determine its stability from the signs of those real parts: if the real
part of each of the eigenvalues is negative at a given equilibrium point,
the solution is stable at that point; if the real part of each eigenvalue
is positive, or if the real part of one eigenvalue is positive and that
of the other is negative, then the solution is unstable at that point.
Finally, if the real part of any of the eigenvalues is zero at a point,
then the point is called nonhyperbolic and its stability in the
neighborhood of that point cannot be determined by this method
\cite{DSref}.  Hence we write
\begin{equation}
	\pmatrix{\xi' \cr \noalign{\medskip}
		 \eta' \cr} =
	\underbrace{
	\pmatrix{2\beta_0(1-3\gamma /2)&-{{2\sigma_0} \over A}(3-3\gamma /2)\cr
		\noalign{\medskip}
		 2\beta_0(1-3\gamma /4) - 2\sigma_0&-{{2\sigma_0} \over A}
			(1-3\gamma /4) - 2\beta_0 \cr}}_{\rm Jacobian}
	\pmatrix{\xi \cr \noalign{\medskip}
		 \eta \cr}
	\quad + \quad \dots \quad ,
\end{equation}
\smallskip\noindent
where $\xi \equiv \beta - \beta_0 \quad {\rm and} \quad \eta \equiv
\sigma - \sigma_0$, and diagonalize the Jacobian case by case.
In this manner we find that for the equilibrium points in the $k=-1$
models under consideration,
$$
	\hbox{if} \quad
	(\beta_0,\sigma_0) = \left\{\matrix{(1,0) \cr \noalign{\smallskip}
					    (-1,0)\cr}\right\},\quad
	\hbox{then}
$$
\smallskip
\begin{equation}
	\left\{\matrix{
		\lambda_1 = 2(1 - 3\gamma /2), \; \lambda_2 = -2 \hfill \cr
		\noalign{\smallskip}
		\lambda_1 = 2(3\gamma /2 - 1), \; \lambda_2 =  2 \hfill \cr}
			\right\}\quad
	\Rightarrow \quad
	\left\{\matrix{
		\hbox{stable,} \; \gamma > 2/3 \hfill \cr
		\noalign{\smallskip}
		\hbox{unstable,} \; \gamma \ne 2/3 \hfill \cr}\right\}.
\end{equation}
\medskip

As expected, the point $(1,0)$ represents a stable solution for
noninflationary, open universes.  The point $(-1,0)$, although an
equilibrium point of the dynamical system, in general represents an
unstable and thus unphysical contracting solution.  Hence a small
perturbation will send the universe along a more rapidly contracting
trajectory as shown in \fig{2a}.  In any case, all models with $\omega
> 0$, negative curvature and noninflationary equations of state must, if
initially expanding, approach equilibrium at $\dot a = 0, \ a \to \infty$
or, if initially contracting, reach a singularity.  If the equation of
state is inflationary, {\it i.e.,\ } if $\gamma < 2/3$, then the
expanding equilibrium solution represented by $(1,0)$ becomes unstable,
and in the range $\omega > 1/2$ we do in fact see inflation for these
models.  Although the existence and stability of the equilibrium points
$(\pm 1,0)$ for $k = -1$ do not depend upon the value of $\omega$, we
again find models which can contract and smoothly reexpand if $-3/2 \le
\omega \le 0$.  This behavior is illustrated in \fig{2b} for models
dominated by pressureless dust.  For $k=-1$ models with $-3/2 \le \omega
\le 1/2$ {\it and} inflationary equations of state, inflation does not
occur despite the fact that the equlibrium solution at $(1,0)$ is
unstable.  This is illustrated in \fig{2c} and \fig{2d}.  In order
to better understand these models, we must move on to an analysis of the
remaining class of equilibrium points.

Let us consider, then, case (ii) of Eq.~\(\ref{EQpts}), {\it i.e.,\ } the
equilibrium points with $\sigma_0 \ne 0$.  Referring back to that
equation, we find that such points will exist only if the condition $k/
\mu(\omega,\gamma) > 0$ is met, which implies that
\begin{equation}
	{4\over 3}\left[1-\left({{2\omega +3}\over {8\omega}}
		\right)^{1/2}\right]
		< \gamma < {4\over 3}\left[1+\left({{2\omega +3}
		\over {8\omega}}\right)^{1/2}\right]
\end{equation}
if $k=-1$, and
\begin{equation}
	\gamma > {4\over 3}\left[1 + \left({{2\omega +3}\over {8\omega}}
		\right)^{1/2} \right]
		\qquad \hbox{or} \qquad
	\gamma < {4\over 3}\left[1 - \left({{2\omega +3}\over {8\omega}}
		\right)^{1/2} \right],
\end{equation}
if $k=+1$.  We can thus think of the curves
$$
	\gamma = {4\over 3}\left[1 \pm \left({{2\omega +3}\over {8\omega}}
		\right)^{1/2} \right]
$$
as bifurcation curves in the parameter space $(\omega, \gamma)$.  By this
one means simply that there are equilibrium points of the dynamical
system which move in or out of existence as those curves are crossed.
However, using Eq.~\(\ref{BDred1}) and Eq.~\(\ref{EQpts}), we find that
these fixed points will lie in the domain $\rho >0$ only if $\gamma <
2/3$, {\it i.e.,\ } only for matter with an inflationary equation of
state.  \fig{3} shows the bifurcation curves in the $\omega$--$\gamma$
plane.  The region between the curves contains new equilibrium points for
$k=-1$; the regions exterior to the curves contains such points for
$k=+1$.  The unshaded region represents that part of the parameter space
with equations of state which can support the solutions represented by
these new fixed points and still maintain $\rho >0$.  It is evident that
the only bifurcation of physical interest occurs for $\gamma =0$, {\it
i.e.,\ } with vacuum energy dominant.  In that case, one finds that two
new equilibrium points appear at $\omega = 1/2$.  If the curvature is
positive $(k=+1)$, then these equilibrium points exist in the regime
$\omega > 1/2$; if the curvature is negative $(k=-1)$, then they exist in
the regime $\omega < 1/2$.  Let us first complete our analysis of the
negative curvature models.  Taking $k=-1$ in Eq.~\(\ref{EQpts}), we have
\begin{equation}
	(\beta_0,\sigma_0) = (\pm(1-3A)^{-1/2},\pm A(1-3A)^{-1/2}) \qquad
		(\omega < 1/2),
\end{equation}
for which we find from the Jacobian that
\begin{equation}
	(\beta_0,\sigma_0) = \left\{\matrix{(+,+) \cr \noalign{\smallskip}
					    (-,-) \cr}\right\} \quad
	\Rightarrow \quad
	\left\{\matrix{\lambda_1 < 0, \quad \lambda_2 < 0 \cr
		\noalign{\smallskip}
                       \lambda_1 > 0, \quad \lambda_2 > 0 \cr}\right\}.
\end{equation}
Here we see that one of the equilibrium points is stable (a sink) while the
other is unstable (a source), as we have already seen explicitly in
\fig{2c} and \fig{2d}.  Let us consider the nature of the stable
solution.  Using the definitions of $\beta$ and $\sigma$, one finds that
the solution at this point is given by
\begin{eqnarray}
		a(t) =&& a_0 + (\Sigma /2)t;	\nonumber	\\
		\phi(t) =&& Ca^2,
\end{eqnarray}
where
\begin{equation}
	\Sigma \equiv (1-3A)^{-1/2}
\end{equation}
and, by assumption, $\rho = \rho_{vac} = \; const$.  Hence, although not
inflationary, the universe will expand forever as we would expect for
a model with negative curvature.  Also, one finds that in order
to satisfy the field equations, the value of $\rho_{vac}$ must be given by
\begin{equation}
	\rho_{vac} = {{3AC}\over{4\pi(3A-1)}} = {{(2\omega+3)C}\over
		{4\pi(1-2\omega)}},
\end{equation}
where $C$ is the integration constant in the the solution for $\phi$,
which we can write as $C = \phi_0 / a_0^2$.  Hence we see that for this
solution, the vacuum energy is completely determined by the initial
conditions of the scale factor and the scalar field, and by $\omega$.

Finally let us consider the models with positive curvature, {\it i.e.,\ }
those with $k=+1$.  The analysis of the equilibrium points is, in light
of our previous discussion, particularly simple.  There are no
equilibrium points in category (i) of Eq.~\(\ref{EQpts}), and the
equilibrium points from category (ii) which lie in the domain $\rho > 0$
exist only for inflationary equations of state and only for $\omega >
1/2$.  Then using Eq.~\(\ref{EQpts}), we find
\begin{equation}
	(\beta_0,\sigma_0) = (\pm(3A-1)^{-1/2},\pm A(3A-1)^{-1/2}) \qquad
		(\omega > 1/2),
\end{equation}
and diagonalization of the Jacobian at those points yields
\begin{equation}
	(\beta_0,\sigma_0) = \left\{\matrix{(+,+) \cr \noalign{\smallskip}
					    (-,-) \cr}\right\} \quad
	\Rightarrow \quad
	\left\{\matrix{\lambda_1 < 0, \quad \lambda_2 > 0 \cr
		\noalign{\smallskip}
                       \lambda_1 > 0, \quad \lambda_2 < 0 \cr}\right\},
\end{equation}
{\it i.e.,\ } both of the equilibrium points are unstable.  So for models
with positive curvature, we have no physically interesting equilibrium
points regardless of the equation of state and the value of $\omega$.
Nevertheless, we may determine the character of the solutions from their
evolution in the $\beta$--$\sigma$ plane.  In general, one expects that
such models should recollapse unless inflation occurs.  Indeed, one finds
that for noninflationary equations of state, {\it i.e.,\ } $\gamma > 2/3,
$ and in the range $\omega > 0$, this is in fact the case.  However, as
before, models with $-3/2 \le \omega \le 0$ and those with $\gamma < 2/3$
have features unique to this theory.  To begin with, one finds that in
the range $-3/2 \le \omega \le 0$, there exist noninflationary models which
do not collapse, despite having positive curvature.  This is illustrated
for pressureless dust in \fig{4a}.  Note that the initial conditions must
be carefully tuned for such models, and in particular that $\sigma < 0$
always, so that $\phi$ is a monotonically decreasing function of time.

For inflationary matter and positive curvature, consider first the range
$-3/2 \le \omega \le 0$.  Here one finds that inflation never occurs,
although there exist models with carefully chosen initial conditions
which expand forever.  All other models, either initially expanding or
initially contracting, collapse to a singularity.  These models are
illustrated in \fig{4b}.  Next, we consider vacuum energy-dominated models
with $k=+1$ but with $0< \omega < 1/2$.  Here we find that inflation
never occurs, and furthermore that {\it all} models, regardless of
initial conditions, collapse to a singularity.  Finally, consider
inflationary models with positive curvature and in the range $\omega >
1/2$, where we have seen that there exist 2 unstable equilibrium points
in the $\beta$--$\sigma$ plane.  Here we find solutions with several
interesting features.  First, in the range $1/2 < \omega < 9/2$,
there are models which start out expanding, and then recollapse rather
than inflate, including some which ``coast'' at $\dot a = 0$ before
beginning their contraction.  If $\omega > 9/2$, then in addition we
find that there exist ``bounce'' models which start out contracting,
but then smoothly reexpand and inflate
--- these are similiar to general relativistic de Sitter spacetime in
$k=+1$ coordinates.
Also in this range of $\omega$,
there are models which are initially contracting, then smoothly
reexpand, and finally recontract until a singularity is reached --- we call
these ``vacillating universes.''  Indeed, there exist models which are
initially expanding, then recontract, and at last smoothly reexpand and
inflate.  If $\omega \ge 9/2$, then in general one finds that the
universe will inflate if and only if its initial conditions lie
in the region bounded by the stable manifolds
of the two equilibrium points.  On the other hand, any model which has
initial conditions which lie between either of the stable manifolds and
the true vacuum in the $\beta$--$\sigma$ plane will collapse back to a
singularity.  The fact that the stable manifolds extend arbitrarily
far back in the negative $\beta$--direction gives rise to
yet another feature unique to this theory:  an arbitrarily rapidly
contracting universe will slow down, smoothly reexpand and inflate if the
energy of the scalar field is chosen just right.  These models represent
a particularly glaring form of ``$\phi$-domination'': whether the universe
collapses to a singularity or inflates depends solely upon the energy of
the scalar relative to the expansion rate.  \fig{4c} illustrates this with
$\omega = 10.5$.  The division of the $\beta$--$\sigma$ plane into
inflationary and noninflationary cells by the stable manifolds is apparent.
\fig{4d} is a closeup of the neighborhood of the equilibrium point in the
expanding regime of \fig{4c}, and one sees explicitly the ``bounce'' and the
``vacillating'' universe models.  The limiting cases of ``vascillation''
are clearly present as well.  For example, a universe which slows down in
its expansion (or contraction), and then speeds back
up may be termed a ``hesitation'' model, whereas one which just stops
but then continues its expansion after a finite time may be
called a ``coasting'' model.  The potentially drastic effects of the
scalar field can also be seen in the special case of $\omega = 9/2$,
where one finds that a necessary condition for inflation to occur is
$\beta > \sigma$, which implies that
\begin{equation}
		{\left(\dot a \over a \right)}_0 > {1\over 2}
		{\left({\dot \phi}\over \phi \right)}_0.
\end{equation}
(However, even if this relation is satisfied, the universe may recollapse
if $\dot\phi / \phi$ is negative enough initially.)  For this value of
$\omega$, a universe which is initially static will begin to expand,
following one of the stable manifolds towards its unstable equilibrium
point, although a small perturbation in the metric or the scalar will
cause either collapse to a singularity or inflation, depending upon
the direction of the perturbation.

\section{Exact Solutions for Flat Space}
\label{exactk0}

Many exact solutions to Eqs.\ (\ref{BDFRW1}-\ref{BDFRW3}) have been found for
the case of flat space ($k = 0$), some of which yield power-laws for the
evolution of the scale factor, the scalar field, and the energy
density\cite{Solns}.  Here we find exact power-law solutions which are
parametrized by the equation of state, and show that these include all
possible power-law solutions to the field equations.

If we define
\begin{equation}
	\lambda \equiv {\sigma \over \beta},
\label{lamdef}
\end{equation}
then using Eqs.~\(\ref{DS1},\,\ref{DS2}) with $k=0$, we can derive
the expression
\begin{equation}
	     	{{d\lambda }\over {d\ln \beta}} =
             {{(\lambda ^2 - A)\left[{3\lambda (1- {\gamma / 2})}
              - {(1 - {{3\gamma }/ 4})}\right]}\over
             {{A(1 - {{3\gamma }/ 2})} -
              {3{\lambda ^2}(1-{\gamma /2})}}}.
\label{k0lamrel}
\end{equation}
Note that any constant value of $\lambda$ which is a root of the cubic
polynomial in the numerator of the right-hand-side will satisfy this equation.
Those roots are
\begin{eqnarray}
	\lambda =&& \pm \sqrt A  	\label{lamvac}			\\
	\lambda =&& {{(1-{3\gamma /4})}\over {3(1-{\gamma /2})}},
\label{lammat}
\end{eqnarray}
the first of which represent the true vacuum solutions, and the last of which
represents a solution with matter (or vacuum energy) present.  Before finding
these solutions explicitly, let us remark that the initial condition
$\dot\phi a^3 \to 0$ as $a\to 0$ is implicit in most of the solutions that
we can generate with this formalism.  To see this, first note that if follows
from the definitions of $\beta$ and $\sigma$ that if $\lambda$ is a constant,
then
\begin{equation}
	\phi = \phi_0 {\left(a\over a_0 \right)}^\alpha,
\label{PLphi}
\end{equation}
where we have set $\phi = \phi_0$ when $a = a_0$, and
\begin{equation}
	\alpha \equiv {\lambda \over {A - {\lambda /2}}}.
\end{equation}
Then we can write
\begin{equation}
 	\dot\phi a^3 \sim \dot a a^{\alpha+2},
\end{equation}
from which we see that if $a \sim t^{C_1}$ at early times, then $\dot\phi
a^3 \to 0$ as $a \to 0$ so long as
\begin{equation}
	C_1 > {1\over{3+\alpha}}.
\end{equation}
One can easily verify that this relation is satisfied for all physically
reasonable perfect fluid equations of state, {\it i.e.,\ } for $\gamma = 0, 1,
\ \hbox{or} \ 4/3$, {\it unless}:
\begin{eqnarray}
	-5/6 <&& \omega < -1/2 \quad \hbox{for} \quad \gamma = 0; \nonumber \\
	-4/3 <&& \omega < -1 \quad \hbox{for } \quad \gamma = 1.
\end{eqnarray}

To see that the $\lambda$ = constant solutions give us all possible
power-law solutions, assume a power-law solution $\phi \propto t^p$, $a
\propto t^q$.  It then follows from the definitions of
$\beta$~and~$\sigma$ that
\begin{equation}
	\beta = {\left({{q+p/2}\over p}\right){\sigma \over A}}
\end{equation}
{\it i.e.,\ } $\lambda$ = constant.

\subsection{True-Vacuum Solutions}

If we set $\lambda = \pm \sqrt A$, we obtain the true-vacuum solutions.
This can be seen by using Eq.~\(\ref{PLphi}) and setting $\rho = 0$ and
$k=0$ in Eqs.\ (\ref{BDFRW1},\,\ref{BDFRW2}), upon which we find that for
$\lambda = \pm \sqrt A$, Eq.~\(\ref{BDFRW1}) is satisfied identically,
whereas Eq.~\(\ref{BDFRW2}) can be integrated to yield the power-law
solutions
\begin{eqnarray}
	a(t) =&& a_0 \bigl[1+(t / t_0)\bigr]^
		{(\pm 2 \sqrt A - 1)/(\pm 6 \sqrt A - 1)},
				\nonumber		    	\\
	\phi (t) =&& \phi_0 \bigl[1+(t / t_0)\bigr]^
		{2/(\pm 6\sqrt A - 1)}			    \label{TVphi}
\end{eqnarray}
where we have set $\phi(t=0) = \phi_0$ and $a(t=0) = a_0$, and $t_0$ is
arbitrary.  These solutions are not valid for $\omega = -4/3$, since the
exponents corresponding to the positive roots in
Eq.~\(\ref{TVphi})
blow up there. The special true-vacuum solution corresponding to
$\omega = -4/3$ was found by O'Hanlon and Tupper in \cite{TV} to be
deSitter in nature, and may be written in the form
\begin{eqnarray}
	a(t) =&& a_0\exp(t/t_0),	\nonumber			\\
	\phi(t) =&& \phi_0 \exp(-3t/t_0).
\end{eqnarray}

\subsection{Solutions with Matter Present}

Now consider the solutions where $\rho \ne 0$.  Then from
$T^{0\nu}{}_{;\nu} = 0$ we have
\begin{eqnarray}
	{{\dot \rho}\over \rho} =&& -3\gamma {\dot a \over a} \nonumber \\
	\rho =&& {\rho _0} {\left(a\over a_0\right)}^{-3\gamma},
\label{RHO1}
\end{eqnarray}
where we have set $\rho = \rho_0$ when $a = a_0$.
Using Eqs\ (\ref{PLphi},\,\ref{RHO1}) in Eqs\ (\ref{BDFRW1},\,\ref{BDFRW2}),
we find that they become
\begin{equation}
	{\left[(1 + {\alpha /2})^2 - A{\alpha ^2}\right]
		{\left(\dot a \over a \right)}^2} =
	{{8\pi \rho_0 {(a / a_0)}^{-3\gamma - \alpha}}
		\over {3\phi_0}}
\label{k0a1}
\end{equation}
and
\begin{equation}
	{{\ddot \phi} + {3\dot \phi {\dot a\over a}}} =
		{{8\pi (4-3\gamma)\rho_0 {(a / a_0)}^{-3\gamma}}
		\over {3 + 2\omega}}.
\label{k0phi1}
\end{equation}
Again setting $a(t=0) = a_0$ and $\phi(t=0)=\phi_0$, Eq.~\(\ref{k0a1})
immediately integrates to
\begin{equation}
	a(t) =	a_0\bigl[1+\chi t\bigr]^{2 / {(\alpha + 3\gamma)}}
\label{k0a2}
\end{equation}
where
\begin{equation}
	\chi \equiv \left({\alpha + 3\gamma}\over 2 \right)
		\left({8\pi \rho_0}\over {3\phi_0 \bigl[(1+\alpha /2)^2 -
		A\alpha ^2\bigr]}\right)
		^{1/2}.
\label{k0chi1}
\end{equation}
Now using Eq.~\(\ref{PLphi}) and Eq.~\(\ref{k0a2}), we see that
\begin{equation}
	\phi(t) = \phi_0\bigl(1+\chi t\bigr)^{{2\alpha} /
		{(\alpha + 3\gamma)}}.
\label{k0phigen}
\end{equation}
Also, using Eq.~\(\ref{RHO1}), we find
\begin{equation}
	\rho(t) = \rho_0\bigl(1+\chi t\bigr)^{{-6\gamma} /
		{(\alpha + 3\gamma)}}
\label{k0rhogen}
\end{equation}
Eqs.\ (\ref{k0a2}-\ref{k0rhogen}) specify all power-law solutions to flat
space FRW/BD cosmology.

Now we can use these results for the various types of matter.  First
consider vacuum energy, where $p=-\rho_{vac}$ is a constant, $\gamma =0,\;
\lambda = {1/3},\; {\rm and} \; \alpha = 4/(2\omega + 1)$.  Then using
Eqs.\ (\ref{k0a2}-\ref{k0rhogen}), we find
\begin{eqnarray}
	a(t) =&& a_0\bigl(1+\chi t\bigr)^{\omega + {1/2}}  \nonumber  \\
	\phi (t) =&& \phi_0\bigl(1+\chi t\bigr)^2
\end{eqnarray}
with
\begin{equation}
	\chi^2_{vac}={{32\pi \rho_{vac}}\over{\phi_0(6\omega+5)(2\omega+3)}}
\end{equation}
in agreement with the known power-law solutions\cite{La,BDinf}.  Note
that if $\omega \le 1/2$, then $\ddot a \le 0$ and inflation does not
occur.  This is noteworthy, as we see that the bare low-energy limit of
string theory, corresponding as it does to a value of $\omega = -1$, does
not permit an inflationary solution for flat space.  In Section IV, we
saw that this result also holds for a universe with curvature.

Next consider a MD era, where
${p=0},\; {\gamma =1}$, ${\lambda ={1/6}},\; {\rm and} \;
{\alpha = 1/(\omega + 1)}$.  We find
\begin{eqnarray}
	a(t) =&& a_0\bigl(1+\chi t\bigr)^{(2\omega +2)/(3\omega +4)}
					\nonumber			\\
	\phi(t) =&& \phi_0\bigl(1+\chi t\bigr)^{2/(3\omega +4)}
					\nonumber			\\
	\rho(t) =&& \rho_0\bigl(1+\chi t\bigr)^{-3(2\omega +2)/(3\omega +4)}
\end{eqnarray}
where
\begin{equation}
	\chi^2_{{\rm MD}} = {{4\pi \rho_0}\over \phi_0}
		\left({3\omega+4}\over{2\omega+3}\right)
\end{equation}
again in agreement with the known power-law solutions\cite{BD,SWein}.

Finally, for a RD era, we have $p={\rho /3}$, $\gamma=4/3,\;
\lambda=0,\; {\rm and} \; \alpha=0$.  From this we find
\begin{eqnarray}
	a(t) =&& a_0\bigl(1+\chi t\bigr)^{1/2}	 \nonumber		\\
	\phi (t) =&& \phi_0\bigl(1+\chi t\bigr)^{-1/2},	 \nonumber	\\
	\rho(t) =&& \rho_0\bigl(1+\chi t\bigr)^{-2}
\end{eqnarray}
where
\begin{equation}
	\chi^2_{{\rm RD}} = {{32\pi \rho_0}\over{3\phi_0}}.
\end{equation}
Note that in this case, the scale factor and energy density evolve just as
they do in ordinary general relativity, independent of the coupling constant
$\omega$.  This is the case because the scalar field has been effectively
decoupled from the other matter due to the vanishing trace of the
stress-energy tensor [cf. Eq.~\(\ref{BDboxeq})].

\section{Conclusions}
\label{concl}

The resurgence of interest in scalar-tensor gravity is justified, in
part, by the formal similarity of the theory to low-energy string theory, and
by the power-law inflationary solution which ameliorates the graceful exit
problem of some general relativistic models of inflation.  However, in order
to create a viable cosmological model with the scalar-tensor theory, one must
understand, to as full an extent as possible, what models are available in the
theory and how those with qualitative differences may be distinguished from
one another.  Here we have shown that there exist a plethora of models which
differ substantively from those present in the standard general relativistic
cosmology.  In general, the models are parametrized by initial conditions of
the scalar field, the value of the BD coupling constant, an initial
value of the Hubble parameter, and an initial density and equation of state of
the matter present.  The first two parameters are not, of course, present in
general relativity.

For flat space, in addition to the standard models which start with a
singularity at the big bang and then asymptotically approach $\dot a = 0$
(or the time-reversal of this behavior) as in general relativity, we have
found ``bounce" models, which pass smoothly from contraction to expansion.
Such models are present for all types of matter in the range of coupling
constant $\omega < 0$.  In addition, we have found that a flat-space
inflationary solution does not exist in this theory in the range $\omega
\le 1/2,$ irrespective of initial conditions.

For a universe with negative curvature, we again find a set of models
which have an analogue in general relativity, in this case those which
start with a singularity and then asymptotically approach the solution
$a \sim t$ (again the time-reversal of this exists with $t \to -t$).
And again, we find ``bounce" models when $\omega < 0$, as well as the
absence of an inflationary solution for $\omega < 1/2$.  We find that in
this latter range of $\omega$, {\it all} expanding negative curvature
models with inflationary-type matter present approach a particular
solution at late times, where $a(t) \sim t$ and $\phi(t) \sim t^2$.

For a universe with positive curvature, the standard models are once more
present, but anomalous behavior occurs both in the range
$-3/2 \le \omega \le 0$ and for inflationary matter.
Again inflation will not occur if
$\omega < 1/2$, in which case all models with vacuum energy dominant
collapse to a singularity only in the range $0 < \omega < 1/2$.  If
$\omega$ is negative, there exist models which perpetually expand for all
types of matter, including vacuum energy.  If vacuum energy is present
and $\omega > 1/2$, we find several classes of models which have no
analogues in general relativity without the addition of a cosmological
constant\cite{Felten}.  First, inflation still does not occur for certain
ranges of initial conditions, those ranges being determined by the
locations of the stable manifolds of the dynamical system in our formalism,
which in turn are determined by the value of $\omega$.  Next we find
``hesitating'' and ``coasting" models, which start with a big bang and
either slow down and ``hesitate'' before continuing their expansion, or
``coast" at a quasistatic value of $a(t)$.  When $\omega > 9/2$,
we find models which expand from a big bang, contract for a time and
then reexpand (``vacillating universes" --- the time-reversed solution also
exists) and models which pass from ``deflation" to inflation,
{\it i.e.,} universes which, although contracting arbitrarily fast, turn
around and ultimately inflate.

Of particular interest are those models with nonzero curvature that are
dominated by vacuum energy which, contrary to the well-accepted
inflationary dogma, do {\it not} approach a corresponding flat-space
solution at late times.  Indeed, we have seen that there exist, for all
values of the coupling constant, vacuum energy-dominated models with
curvature for which there
are no corresponding flat-space solutions at all.  In such cases, the
initial energy of the scalar field relative to the expansion rate
sometimes plays a key role in determining whether inflation will occur, a
fact which appears to have been largely overlooked in the literature.

\section{Acknowledgements}
This research was supported in part by the National Science Foundation under
Grants No.~PHY89-04035 and PHY90-08502.  We are grateful to J.~Barrow and
D.~Wands for helpful communications.

\gdef\journal#1, #2, #3, 1#4#5#6{		% Journal reference. Comma sets
    {\sl #1~}{\bf #2} (1#4#5#6), #3}		% off: name, vol, page, year
\def\pr{\journal Phys. Rev., }
\def\prd{\journal Phys. Rev. D, }
\def\prl{\journal Phys. Rev. Lett., }
\def\rmp{\journal Rev. Mod. Phys., }
\def\np{\journal Nucl. Phys., }
\def\npb{\journal Nucl. Phys. B, }
\def\pl{\journal Phys. Lett., }
\def\plb{\journal Phys. Lett. B, }
\def\apj{\journal Astrophys. Jour., }
\def\apjl{\journal Astrophys. Jour. Lett., }
\def\annp{\journal Ann. Phys. (N.Y.), }

\begin{figure}
\caption{The evolution of solutions in the $\beta$--$\sigma$ plane for
	flat space.
	(a) Matter is $\gamma=1$ (pressureless dust);
	and $\omega = 9/2$, representative of $\omega > 0$.
	Solutions to the right of the line $\beta = \sigma /2$
	represent expanding universes; those to the left are contracting
	universes.  The shaded regions require negative energy density
	and so are disallowed physically.
	(b) Same as (a), except $\omega = -1/2$.
	The line ${da\over dt}=0$ has moved into the regime of
	positive energy density, so that some models pass
	smoothly from contraction to expansion, a general feature
	of models with $-3/2 \le \omega \le 0$.
	(c) Flat space $\gamma=0$ (vacuum energy dominated) models, with
	$\omega = 9/2$.  All initially expanding universes become inflationary
	--- they flow out along the prominent channel to the left ---
	regardless of initial conditions when $k=0$ and $\omega > 1/2$.
	(d) Flat space, $\gamma=0$ (vacuum energy dominated) models with
        $\omega = -1/2$, representative of $-3/2 \le \omega \le 0$.  Inflation
	does not occur, although all initially expanding universes
	continue to expand forever.
\label{fig1}}
\end{figure}

\begin{figure}
\caption{Some models with $k=-1$ (negative curvature).
	(a) $\gamma=1$ (pressureless dust) with $\omega > 0$.
	(b) $\gamma=1$ with $\omega < 0$.
	(c) $\gamma=0$ (vacuum energy) with $\omega < 0$.
	(d) a closeup of the stable equilibrium point in (c).
	In (a) $\omega = 9/2$; in (b-d) $\omega = -1/2$.
\label{fig2}}
\end{figure}

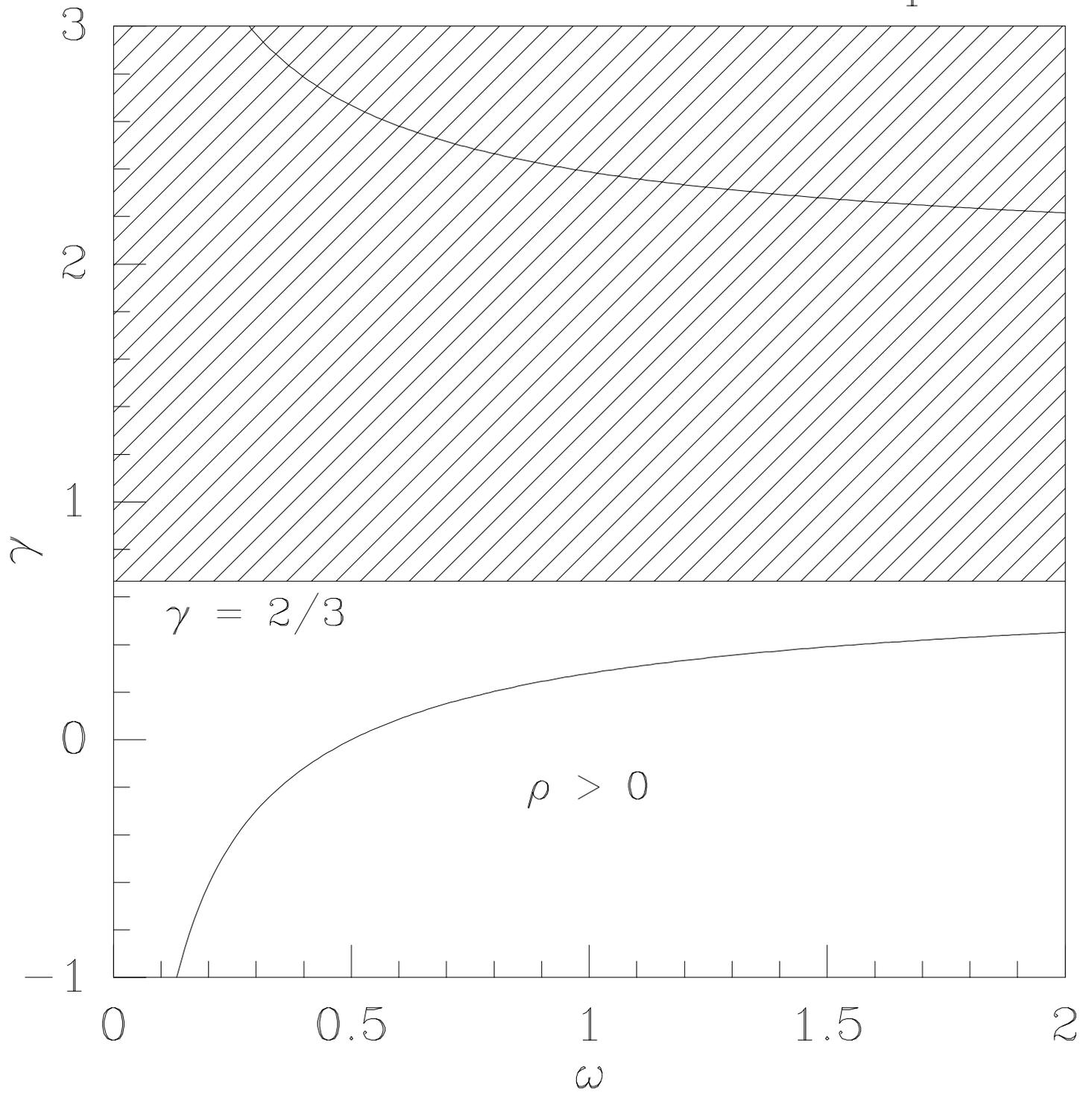
\begin{figure}
\caption{Bifurcation diagram in the parameter space $\omega$--$\gamma$.
	The shaded region is disallowed physically as it requires
	$\rho < 0$.  The regions exterior to the curves contain
	bifurcated equilibrium points for $k=+1$; the region interior
	to the curves contains bifurcated equilibrium points for $k=-1$.
	The bifurcation for $k=+1$ and $\gamma = 0$ (vacuum energy)
	is physically permissible and occurs when $\omega = 1/2$.
\label{fig3}}
\end{figure}

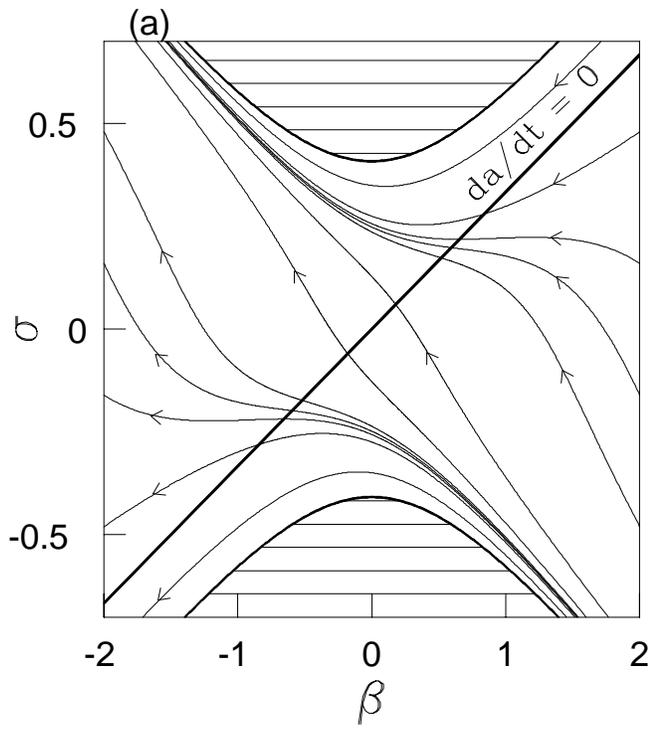
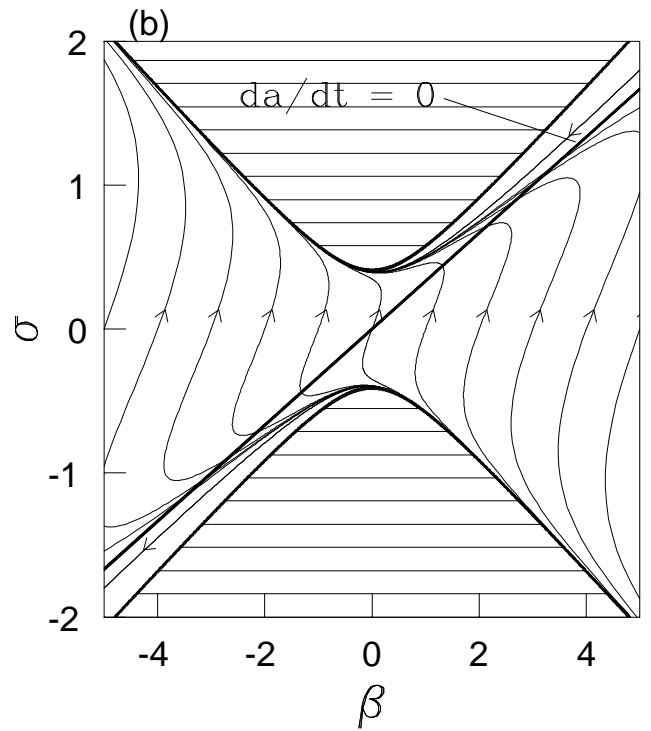
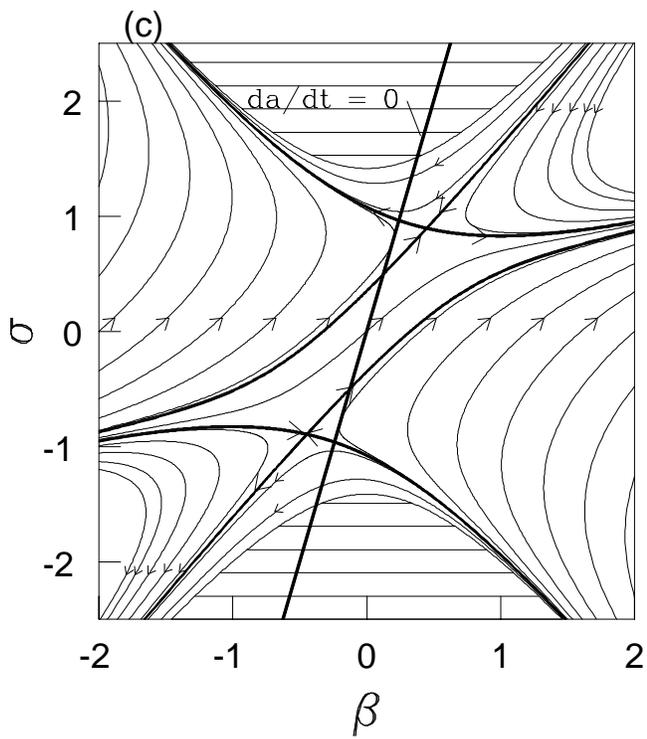
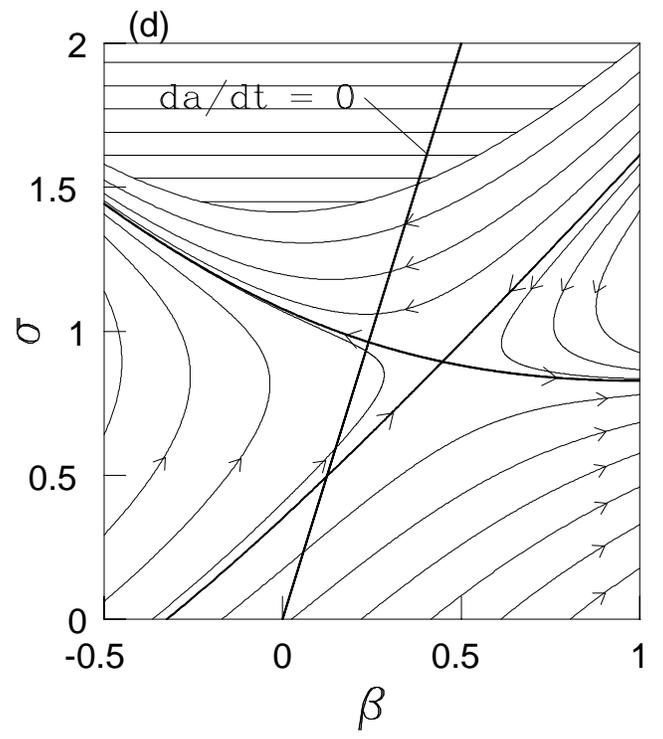
\begin{figure}
\caption{Some models with $k=+1$ (positive curvature).
	(a) $\gamma=1$ (pressureless dust) and $\omega=-1/2$, representative
	of $-3/2<\omega<0$.  Some initially expanding models do not
	recollapse, despite positive curvature.
	(b) $\gamma=0$ (vacuum energy dominant); and $\omega=-1/2$,
	representative of $-3/2<\omega<0$.
	Most expanding solutions recollapse; a few expand
	forever, but inflation never occurs.
	(c) $\gamma=0$ (vacuum energy dominant); and $\omega = 10.5$,
	representative of $\omega > 9/2$.
	Most expanding universes inflate, but some do not.  Here an
	arbitrarily rapidly {\it contracting}
	universe will slow down, stop and inflate if the energy
	of the scalar relative to the expansion rate is carefully
	chosen.
	(d) A closeup of an unstable equilibrium point in (c).
	The stable manifold separates models which inflate from those which
	recollapse.  Also apparent are ``vacillation" universes which start
	out contracting, then briefly expand, and finally recontract
	to a singularity.
\label{fig4}}
\end{figure}

\end{document}